\documentstyle[11pt,aaspp4,flushrt]{article}

\newcommand{\G}{\Gamma}
\newcommand{\g}{\gamma}
\newcommand{\ext}{{\rm ext}}
\newcommand{\peak}{{\rm peak}}
\newcommand{\Min}{{\rm min}}
\newcommand{\tobs}{{t_{\rm obs}}}

\begin{document}

\title{Energy-Dependent Gamma-Ray Burst Peak Durations and
Blast-Wave Deceleration}

\author{James Chiang\altaffilmark{1}}

\affil{E.\ O.\ Hulburt Center for Space Research, Code 7653, Naval 
Research Laboratory, Washington DC 20375-5352}
\altaffiltext{1}{NAS/NRC Research Associate}

\begin{abstract}
Temporal analyses of the prompt gamma-ray and X-ray light curves of
gamma-ray bursts reveal a tendency for the burst pulse time scales to
increase with decreasing energy.  For an ensemble of BATSE bursts,
Fenimore et al.\ (1995) show that the energy dependence of burst peak
durations can be represented by $\Delta t \propto E^{-\gamma}$ with
$\gamma \simeq 0.4$--$0.45$.  This power-law dependence has led to the
suggestion that this effect is due to radiative processes, most
notably synchrotron cooling of the non-thermal particles which produce
the radiation.  Here we show that a similar power-law dependence
occurs, under certain assumptions, in the context of the blast-wave
model and is a consequence of the deceleration of the blast-wave.
This effect will obtain whether or not synchrotron cooling is
important, but different degrees of cooling will cause variations in
the energy dependence of the peak durations.
\end{abstract}


\section{Introduction}

Since their discovery in the late 1960s, gamma-ray bursts (GRBs) have 
remained enigmatic despite the fact that thousands of bursts have been 
detected by various instruments over the last $\sim 30$ years.  The 
recent X-ray, optical and radio afterglow observations of bursts 
detected by the BeppoSAX satellite have enabled significant advances 
in our understanding of these objects.  Redshift measurements 
associated with the afterglows of bursts GRB~970508 (Metzger et al.\ 
1997), GRB~971214 (Kulkarni et al.\ 1998), and GRB~980425 (Tinney et 
al.\ 1998) provide convincing evidence that GRBs are extragalactic and 
may be as distant as $z = 3.4$ or as nearby as $z = 0.0085$.  The 
temporal decay of the afterglow emission is roughly consistent with 
the simplest fireball/blast-wave interpretations (Wijers, Rees \& 
M\'esz\'aros 1997; Waxman 1997), and these models provide a specific 
theoretical context in which to understand the wealth of burst data 
which existed prior to BeppoSAX and which is still largely 
unexplained.

It has been argued that the spectral shapes of a substantial number of
bursts, particularly those which have detailed spectra from the Burst
and Transient Source Experiment (BATSE) and the other {\it Compton
Observatory}\ instruments, are due to synchrotron emission from a
shock accelerated distribution of non-thermal electrons (e.g., Tavani
1996).  Furthermore, the spectra of X-ray, optical and radio afterglow
emission also appear to be due to synchrotron radiation and exhibit
characteristic signatures such as power-law behavior consistent with
optically thin synchrotron emission in the optical and X-ray bands
(Djorgovski et al.\ 1997; Frontera et al.\ 1998; Galama et al.\ 1998),
self-absorption in the radio band (Katz \& Piran 1997; Frail et al.\
1997), and spectral index changes of $\Delta \alpha = 0.5$ in the
optical, indicative of synchrotron cooling (Galama et al.\ 1998).

However, even in the context of a specific dynamical and emission
model, the varied complexity of the spectral and temporal properties
of prompt GRB light curves remains an outstanding problem.  In this
work, we address one aspect of this problem: the tendency, for a given
GRB, for longer burst peak durations at lower observed energy bands.
We examine this issue in terms of the blast wave model and demonstrate
how a similar tendency arises due to blast wave deceleration and how
different degrees of synchrotron cooling can modify the basic effect
to produce a range of behavior.  In the remainder of this paper, we
give a brief summary of the relevant aspects of the observed energy
dependence of burst light curves (\S~2), describe the dynamics and
emission properties of the basic blast-wave model and how they relate
to this effect (\S~3), and finally, discuss some of the strengths and
weaknesses of this interpretation and suggest some avenues for further
investigation (\S~4).

\section{Energy Dependence of Burst Light Curves}

The general behavior of burst light curve shapes was noted by Fishman
et al.\ (1992).  The individual pulses tend to be sharper at higher
energies and are often characterized by fast rise times and slow
decays, the so-called FREDs (fast rise, exponential decay).  Link,
Epstein \& Priedhorsky (1993) put these descriptions on a quantitative
basis by applying autocorrelation functions and skewness analyses to
individual burst light curves.  The average properties of burst peak
durations as a function of energy were studied by Fenimore et al.\
(1995) for a collection of 45 bright BATSE bursts.  They found that
the average peak width can be described by $\Delta t \propto
E^{-0.45}$.  Furthermore, they found that the energy-dependent
autocorrelation function possesses a nearly universal shape with the
width of the autocorrelation function having this energy scaling.

From studying Ginga bursts, Fenimore (1998) also notes that there is a
tendency for precursor emission or ``pre-activity'' in the lower
energy X-ray bands prior to the main peak at higher energies.  He
suggests that the physical processes responsible for the longer burst
durations at lower energies also account for this pre-activity.
Furthermore, Fenimore finds that the location of the light curve peaks
are not substantially delayed at lower energies versus high.  Fenimore
claims that these phenomena appear to be inconsistent with models
which only consider synchrotron cooling of injected electrons since
they predict longer time scales at lower energies only for the
decaying part of the light curve and also that the light curve peaks
at the lowest X-ray energies should occur only after the high energy
emission has died away (e.g., Kazanas, Titarchuk \& Hua 1998).
However, Dermer (1998) shows that if the electrons are injected at a
constant rate over a finite period of time, then over that time
period, the pulse widths will indeed vary with energy due to
synchrotron cooling, but the initial rise of each of the pulses will
essentially line up.  In this paper, we consider the somewhat more
complex situation in which the electron injection is governed by the
deceleration of the blast wave and the conversion of its bulk kinetic
energy to non-thermal particle energy and magnetic fields.

\section{Blast-Wave Deceleration and the Synchrotron Spectrum}

Several descriptions of the physics of the blast wave model already
exist in the literature and we refer the reader to Panaitescu \&
M\'esz\'aros (1998) and Chiang \& Dermer (1998) and the references
therein for further details.  The model consists of an expanding shell
of material with a large initial bulk Lorentz factor, $\G \sim
10^2$--$10^3$. This shell decelerates as it interacts with an external
medium.  A shock forms, and in the shocked material, magnetic fields
are generated, and particles are accelerated.  An initial phase of
nearly free expansion ends when the expanding blast wave shell begins
to sweep up ambient matter at a rate sufficient to decelerate it
significantly.  During this deceleration phase, the bulk Lorentz
factor obeys $\G \propto r^{-\zeta}$ where $r$ is the radius of the
shell with $\zeta = 3/2$ for non-radiative (or adiabatic) expansion
and $\zeta \simeq 3$ for radiative expansion.

The rate at which the bulk kinetic energy of the blast wave shell is
converted to internal energy is given by
\begin{equation}
\frac{dE}{dt} = c^3 A(r) m_p n_\ext(r) \beta\G(\G - 1)
\label{dEdt}
\end{equation}
where $A(r)$ is the area of the blast wave shell as a function of
radius and equals $4\pi r^2$ for a spherically expanding shell,
$n_\ext$ is the ambient particle number density, and $\beta = (1 -
1/\G^2)^{1/2}$ (Blandford \& McKee 1976).  In order to perform
concrete calculations, it is conventionally assumed that the magnetic
field energy density and non-thermal electron energies can be
described by parameters $\xi_B$ and $\xi_e$ which measure the degree
to which these components are in equipartition with the swept-up
protons.  In the co-moving frame of the shell, these protons have
energies $\G m_p c^2$.

At the early stages of the deceleration, when the bulk Lorentz factor
is $\G \ga 10^2$, most of the emission seen by an observer comes from
the nearest part of the blast wave shell within a narrow cone with
opening angle $\sim 1/\G$.  Hence, the emission properties during
these stages are reasonably well accounted for by the state of the
material along the line-of-sight connecting the observer to the center
of the explosion.  We model the injected non-thermal electrons as a
power-law distribution with a sharp lower-energy cut-off, $dN/d\g
\propto \g^{-p}$ for $\g > \g_\Min$, all radiating from within a
region of uniform magnetic field energy density.  Fits to burst
spectra indicate that the power-law index is typically very steep with
$p \ga 5$ (Tavani 1996).  The minimum electron Lorentz factor (and
hence the characteristic electron energy) is related to the
equipartition parameter by $\g_\Min = \xi_e (m_p/m_e) \G$.  It follows
then that the peak energy of the $\nu L_\nu$ synchrotron spectrum of
this injected electron distribution is given by (Katz \& Piran 1997;
Chiang \& Dermer 1998)
\begin{eqnarray}
E_\peak &\sim & 3\times10^{-8} m_e c^2 n_\ext^{1/2} \xi_B^{1/2}
                \xi_e^2 \G^4(r) \\ 
        &\propto & \tobs^{-4\zeta/(2\zeta+1)},
\label{epeak}
\end{eqnarray}
where $\tobs$ is the elapsed time as seen by the observer, and we have
assumed that the burst is nearby (i.e., $z \simeq 0$).  We also have
used the relation $\tobs \propto r^{2\zeta + 1}$ which holds during
the deceleration phase of the blast wave when $\G \propto r^{-\zeta}$.
For the discussion which follows, we will further assume that $\xi_B$,
$\xi_e$, and $n_\ext$ are constants unless otherwise indicated.

If synchrotron cooling and other energy loss mechanisms are not
important, then the instantaneous $\nu L_\nu$ synchrotron spectrum
near $E_\peak$ during the prompt burst phase can be described
approximately by a broken power-law (Chiang \& Dermer 1998):
\begin{eqnarray}
\nu L_\nu &\propto & r^2 \G^4(r) (E/E_\peak)^{\lambda_\pm}\\
          &\propto & \tobs^{(2+4\zeta(\lambda_\pm-1))/(2\zeta+1)},
\label{nuLnu_t_dependence}
\end{eqnarray}
where $\lambda_+ = (3-p)/2$ for $E > E_\peak$, $\lambda_- = 4/3$ for
$E < E_\peak$ (Rybicki \& Lightman 1979; Katz 1994), and in
eq.~\ref{nuLnu_t_dependence}, we explicitly write out the
$\tobs$-dependence of $\nu L_\nu$ at a given energy.  We note that
this time dependence only holds during the deceleration phase of the
blast wave and only for energies less than the initial value of
$E_{\peak}$ ($= E_\peak(\G_0)$ where $\G_0$ is the initial blast wave
bulk Lorentz factor).  The remarkably sharp peaks seen in the $\nu
F_\nu$ spectra of a number of bursts are consistent with this spectrum
(Tavani 1996; Schaefer et al.\ 1998).  If synchrotron cooling is
important, then the resulting spectrum will be somewhat more complex.
In this case, $\lambda_+ = (2-p)/2$, describing the softer spectrum of
a cooled power-law electron distribution, and the $\nu L_\nu$ peak
will be much broader and flatter, and at energies below the peak, the
index will be $\lambda_- \simeq 1/2$ rolling over to $4/3$ at still
lower energies (Sari, Piran \& Narayan 1998; Chiang \& Dermer 1998).

Tavani (1996) and Schaefer et al.\ (1998) have examined the spectra of
a number of individual bursts and find them to be consistent with an
uncooled distribution of electrons (i.e., $\lambda_- = 4/3$).
However, Schaefer et al.\ also analyze the composite burst spectrum
constructed from a sample of 19 bright BATSE bursts, including those
bursts which they also examined individually.  They find an
``average'' lower energy spectral index of $\lambda_- = 0.77$.
Although some of the features of the composite spectrum can be
attributed to the range of values for the peak energy among the
different bursts, it seems likely that this softer index is due to
some of the co-added burst spectra having been significantly affected
by synchrotron cooling.

In order to demonstrate the effect of blast wave deceleration, we
consider the case of uncooled electron distributions.  Using values
$\zeta = 3/2$ and $p = 6$, we have from eq.~\ref{nuLnu_t_dependence}
\begin{eqnarray}
\nu L_\nu &\propto& \tobs, \qquad E < E_\peak \label{nuLnu_timing} \\
          &\propto& \tobs^{-13/4}, \qquad E > E_\peak, \nonumber
\end{eqnarray}
and we see that the declining portion of a blast wave pulse will only
occur after $E_\peak$ has passed through the energy band at which the
observations are taking place.  Thus, there will be an energy-dependent
delay for the light curve decline given by inverting the expression
for the energy of the synchrotron peak (eq.~\ref{epeak}):
\begin{eqnarray}
\tobs &\propto& E^{-(2\zeta+1)/4\zeta} \label{tobs_eqn} \\
      &\propto& E^{-0.67}, \qquad \zeta = 3/2 \nonumber \\ &\propto&
      E^{-0.58}, \qquad \zeta = 3 \nonumber.
\end{eqnarray}
Although this analysis predicts that the timing of the pulse peaks at
lower energies will also be displaced, in practice the intrinsic
curvature of the synchrotron spectrum at $E_\peak$ will ameliorate
this discrepancy (cf. Fig.~1).

Of course, synchrotron cooling will always be present in this model,
though perhaps to varying degrees.  Using our blast wave code (Chiang
\& Dermer 1998), we now present detailed calculations for cases of
relatively weak ($\xi_B = 10^{-6}$, $\G_0 = 300$) and strong ($\xi_B =
10^{-2}$, $\G_0 = 100$) magnetic fields.  The initial bulk Lorentz
factor, $\G_0$, has been adjusted in each case so that the initial
value of $E_\peak$ is approximately the same, and the fraction of the
total kinetic energy (eq.~\ref{dEdt}) made available in the form of
non-thermal electrons for both cases is $10$\%.  This latter condition
ensures that the bulk motion of the blast wave shell in both
calculations is well described by non-radiative deceleration ($\zeta
\simeq 3/2$) and that any relevant differences will be solely due to
the effects of the different magnetic fields.

Figure~1 shows the results for the weak magnetic field case.  In the
upper panel, we plot the $\nu L_\nu$ spectra at times $\tobs =
10^{\{-1,-0.5,0,\ldots\}}$ seconds after the initial fireball event.
The vertical dotted lines indicate energies $25, 57, 115,
320$~keV---the lower bounds of the four standard BATSE LAD channels
(cf.\ Fenimore et al.\ 1995).  In the lower panel, the light curves at
these four energies are shown, each normalized to unity at their
respective maxima.  We see that this simple model reproduces a single
FRED-like pulse quite naturally.  In addition, the light curve decline
is delayed at lower energies and the energy dependence of the
half-maximum pulse widths is $\Delta t \propto E^{-0.66}$ consistent
with eq.~\ref{tobs_eqn} for a non-radiative blast wave with no
cooling.  The lower energy peaks are also delayed, but not excessively
so.  We note, however, that the pulses are less sharply peaked at
these lower energies.  In Figure~2, we show the results for a strong
magnetic field.  As we noted above, the spectral peaks are much
flatter and broader, and the lower energy index is close to $\lambda_-
= 1/2$.  The effect on the light curves is also substantial.  The
light curve peaks are broader, but in addition, the energy dependence
of the half-maximum pulse widths is significantly weaker with $\Delta
t \propto E^{-0.39}$.  These shorter delays for the onset of the light
curve declines are due to the combination of the blast wave
deceleration and synchrotron cooling both pushing the $\nu L_\nu$ peak
down in energy faster than either process would do so individually.

\section{Discussion}

Given our assumptions, the effect of pulse broadening at lower
energies will obtain for any blast wave deceleration model, whether
the prompt burst emission is due to internal shocks or external
shocks, as we have described here.  Furthermore, the energy dependence
of the pulse widths will vary depending on the degree of synchrotron
cooling, and we find that the range of this energy dependence
($E^{-0.4}$--$E^{-0.66}$) is suggestively close to the results found
by Fenimore et al.\ (1995).  This simple picture, however, cannot
explain the pre-activity or precursor behavior described by Fenimore
(1998).  In addition, especially for strong cooling, the pulse shapes
will be less sharply peaked at lower energies, which implies that the
autocorrelation function may not have the universal shape described by
Fenimore et al.\ (1995).

A substantially more complex blast wave model than the one we have
presented here is certainly required to describe GRBs, and any
additional complexities may either mitigate or worsen the above
discrepancies with the observations.  At a minimum, the electron and
magnetic field equipartition parameters are not constant throughout
the blast wave evolution as we have assumed.  As the blast wave
evolves, the expansion time scale also grows with increasing radius.
There would then be more time for equipartition to obtain, and one
might expect the electron and magnetic field components to come closer
to true equipartition with the swept-up protons at later times.  Such
an effect may account for the softer pre-activity phase since
$E_\peak$ will increase as $\xi_e$ and $\xi_B$ increase until the
temporal dependence implied by the deceleration and synchrotron losses
begin to dominate and the effects we have described will then control
the light curve evolution.  The effects of evolving equipartition
parameters in more complex blast wave model calculations can certainly
be studied, but one may suspect that only rather contrived electron
energy and magnetic field evolution will reproduce the behavior
described by Fenimore (1998).  In this case, the ultimate answers may
lie in doing the microphysics: magnetic field generation and particle
acceleration, in other words, turbulent relativistic
magneto-hydrodynamics.

On the observational side, at least three additional investigations
should be conducted in order to help guide the theoretical modeling.
First, a cross-correlation or similar analysis of the prompt burst
light curves in the various energy bands should be performed in order
to quantify the energy-dependent delay of the pulse peaks.  Second, in
addition to computing properties averaged over an ensemble of bursts,
the {\em distributions} of the energy dependences of the pulse widths
and peak delays should also be computed.  This information will
determine the range of parameter space for which any given model must
account.  Finally, insofar as signal-to-noise limitations prevent
time-resolved spectra, the spectral properties integrated over the
prompt burst phase should be analyzed for each individual burst.
Specifically, if there is evidence for synchrotron cooling in the
spectrum, which will be signified by a $\nu^{1/2}$-dependence in the
energy bands of interest below the $\nu F_\nu$ peak, then synchrotron
cooling may well be important, and in the context of the blast wave
model, such spectral features should be correlated with the energy
dependence of the pulse durations.

\acknowledgements
The author acknowledges helpful discussions with C.\ D.\ Dermer.  This
work was performed while J.C. held a National Research Council-NRL
Research Associateship.

\newpage
\begin{figure}
\epsfysize=6.75in
\centerline{\epsfbox{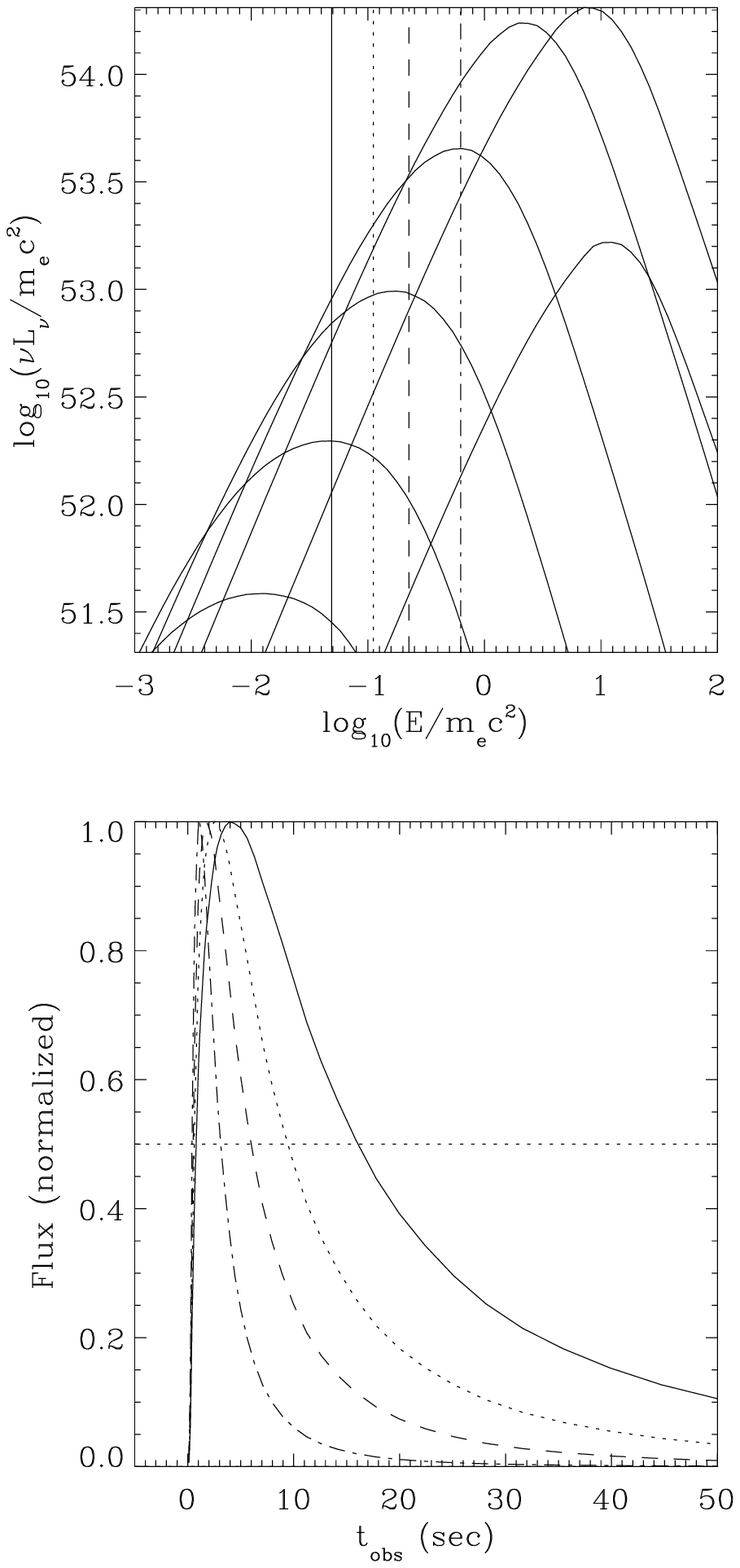}}
\caption{Spectral and light curve calculations in the blast wave model
for weak magnetic fields.  The initial bulk Lorentz factor of the
blast wave shell is $\G_0 = 300$, and the magnetic field equipartition
parameter is $\xi_B = 10^{-6}$; other model parameters are $\xi_e =
1$, $n_\ext = 10^2$~cm$^{-3}$, $p = 6$.  {\em Upper panel:}\ Spectra
for different observer times $t = 10^{\{-1,-0.5,0,\ldots\}}$~s.  The
vertical lines denote the lower energy bounds for the four BATSE LAD
channels: 25 (solid), 57 (dotted), 115 (dashed), and 320~keV
(dot-dashed).  {\em Lower panel:}\ Light curves for each of the above
energies.  The energy dependence of the pulse width at half-maximum is
$\Delta t \propto E^{-0.66}$.}
\label{pw_plot_1}
\end{figure}

\begin{figure}
\epsfysize=6.75in
\centerline{\epsfbox{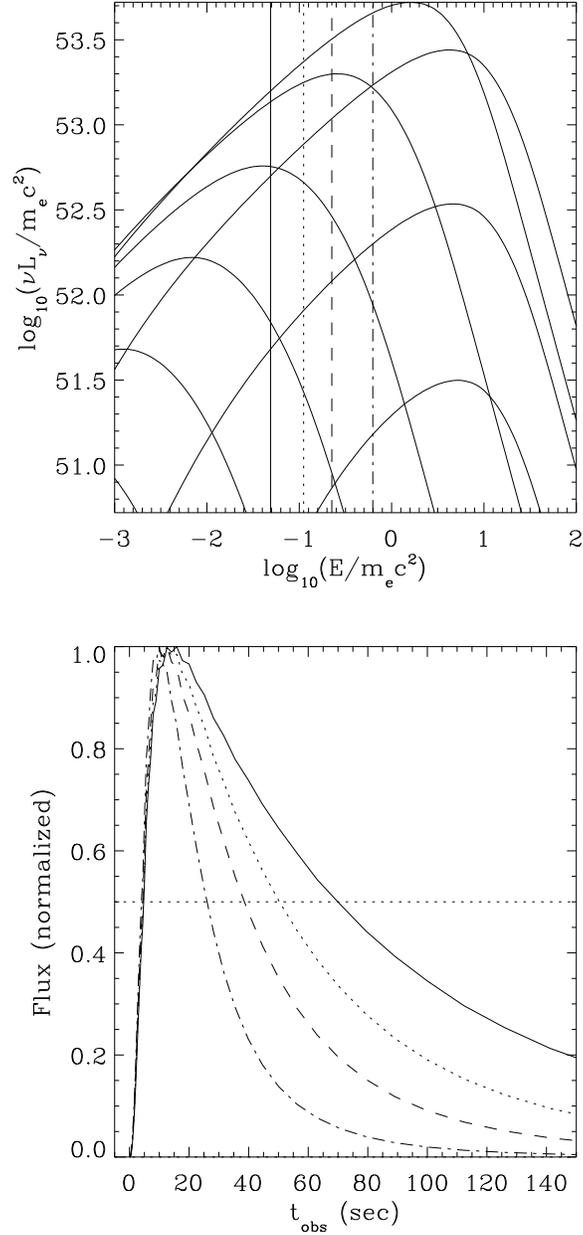}}
\caption{Strong magnetic field case.  Here $\xi_B = 10^{-2}$ and 
$\G_0 = 100$; the other parameters are the same as for Fig.~1.  {\em
Upper panel:}\ Spectra for different observer times $t =
10^{\{-1,-0.5,0,\ldots\}}$~s. {\em Lower panel:}\ Light curves for the
four BATSE LAD lower energy bounds.  The energy dependence of the
pulse width at half-maximum is $\Delta t \propto E^{-0.39}$.}
\label{pw_plot_2}
\end{figure}

\end{document}